\documentclass[12pt]{iopart}

\usepackage{iopams}  
\usepackage{graphicx}
\begin{document}

\title[Wave transformation -- Simulation and Application]{Wave transformation -- Simulation and Application}

\author{Yinpeng Wang}
\address{37th Xueyuan Road, Beihang University, Beijing, 100191, China}
\ead{16081072@buaa.edu.cn}

\vspace{10pt}
\begin{indented}
\item[] \today
\end{indented}

\begin{abstract}
This paper focus on the theoretical analysis and simulation of electromagnetic wave transforms, which is widely encountered in teaching physics. When the electromagnetic wave is not consistent with the shape of the object, it is often necessary to introduce electromagnetic wave transforms to analyze the interaction between the two. In this paper, we discuss three common waves, namely plane wave, cylindrical wave and spherical wave. We first analyze the mathematical transformation equations between them, and then use computer programs to draw a simulation image to intuitively show the derivation results.  During the derivation, we corrected an error in a classic electromagnetic textbook. Finally, several specific electromagnetic examples are solved by electromagnetic wave transformation. These examples include the scattering of plane waves by ideal conductor cylinder, sphere and ideal dielectric cylinder. The simulation results are displayed intuitively in the form of images.
\end{abstract}

%
%
%
%
%


\section{Introduction}
Electromagnetic scattering is an important subject in the field of electromagnetics \cite{Qi,Li}. When the electromagnetic wave propagates to the interface between two media, refraction and reflection will occur. When the geometric shape of the isophase plane of the electromagnetic wave is consistent with the boundary, the electromagnetic field on both sides of the interface can be easily solved (for example, when the incident wave is a plane wave and the interface is infinite plane or the incident wave is spherical wave and the interface is spherical). When the shape of the isophase plane of electromagnetic wave is not consistent with the shape of the boundary, it is necessary to use one form of wave function to adapt to another form of boundary conditions. It is often complicated to solve such problems directly. A common idea is to decompose the incident electromagnetic wave into a group of electromagnetic waves which satisfy the geometric boundary conditions \cite{1}. The superposition coefficient is determined by the boundary conditions, and the field distribution is solved.

Harrington elaborated the fundamental principle of wave transformation in his book \cite{2} and gave some applications. However, limited by age, the illustrations in the book are obsolete and therefore not intuitive enough for readers. With the widespread application of computer simulation in teaching, physical laws can be exhibited in a more vivid form. In this paper, we mainly discuss how to visually demonstrate the wave transformation process by computer simulation. Herein, a series of simulations are carried out for several cases on wave transformation in Harrington's works. We suppose it can aid undergraduate students intuitively understand the process of electromagnetic wave transformation and stimulate their enthusiasm for learning. In addition, this paper points out an improper deduction in Harrington's book and corrects it in the appendix.

This paper organizes into the following parts: in part 2, we expound on the basic principle behind the wave transformation and derive the formula. Then, we depict the outcome of computer simulation. In part 3, we illustrate the corresponding application of practical electromagnetic wave transformation. In part 4, we summarize our work. In addition, we correct an improper formula in Harrington's works in the appendix.

\section{Formulation and Simulation}
Common orthogonal curvilinear coordinate systems include rectangular, cylindrical and spherical coordinate system. The corresponding waves are plane wave, cylindrical wave and spherical wave. In this paper, we will exhibit the analytical formula of the plane wave decomposition into superposition of cylindrical waves and spherical waves, and the cylindrical wave decomposition into that of spherical waves. The corresponding computer simulation pattern are also demonstrated, helping students better understand this process.
\subsection{Plane Wave to Cylindrical Wave}
For a typical plane wave, by selecting a suitable coordinate system, the wave function can be expressed as $Ae^{ikx}$. For a typical cylindrical wave, it can be written as $AJ_n\left(k\rho\right)e^{in\phi}$, where $J_n(\cdot)$ is a Bessel function of order $n$. As the family of Bessel functions is orthogonal complete, it is feasible to implement transform on the plane wave into the cylindrical wave \cite{3}.

\begin{equation}
e^{ikx}=e^{-ik\rho \cos \theta}=\sum_{n=-\infty}^{+\infty}a_{n}J_{n} (k\rho)e^{in\phi},
\end{equation}
where the coefficient $a_n$ can be obtainted by the Laurent expansion of the complex function $f(z)=e^{\frac{z}{2}(t-\frac{1}{t})}$ in the interval $0<\left|z\right|<+\infty$.

\begin{equation}
f(z)=e^{\frac{z}{2}(t-\frac{1}{t})}=\sum_{n=-\infty}^{+\infty}{J_n\left(z\right)t^n},0<\left|z\right|<+\infty
\end{equation}

Let $t=-ie^{i\theta},\ z=k\rho$, we have

\begin{equation}
e^{ikx}=e^{-ik\rho \cos \theta}=\sum_{n=-\infty}^{+\infty}i^{n}J_{n} (k\rho)e^{in\phi}.
\end{equation}

\begin{figure}[htbp]
\centering
\includegraphics[scale=.65]{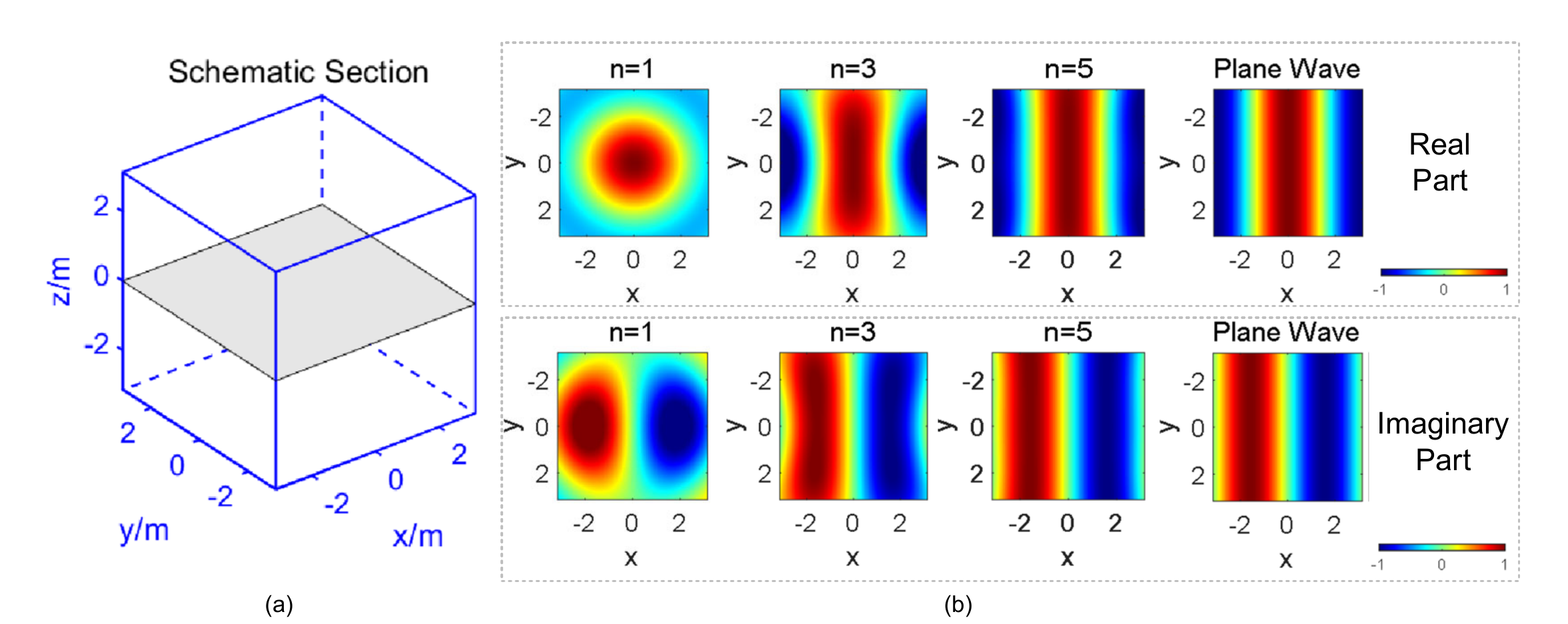}
\caption{Plane wave decompose into cylindrical wave. (a) Schematic section of the observation (b) Comparison of superposition of cylindrical wave and plane wave when the truncation n is 1,3,5 respectively.}
\label{Fig1}
\end{figure}

Here we demonstrate the computer simulation pattern, which on the one hand verifies the correctness of the derived formula, and on the other hand clearly illustrates the process of the plane wave decomposition. Suppose the plane wave propagates along the $+x$ direction, it can be expressed as $e^{ikx}$. Fig. \ref{Fig1}(a) explains the observation section while (b) reveal the superposition of cylindrical waves and plane waves when the number of truncation terms $n$ are 1, 3 and 5 respectively.

The upper and lower groups emerge the fields of real and imaginary parts respectively. With the increase of the number of truncation terms $n$, the superposition of cylindrical waves is closer to the plane waves. In fact, when $n=5$, they can be approximately equal.

\subsection{Plane Wave to Spherical Wave}
For the transformation of plane wave to spherical wave, it is more common to assume that the wave propagates along $+z$ direction. Therefore, it can be expressed as $Ae^{ikz}$ and the spherical wave can be written as $Ac_n\left(kr\right)P_n\left(cos\theta\right)$, where $P_n\left(cos\theta\right)$ is a Legendre polynomial of order $n$. As the family of Legendre polynomials is orthogonal complete, it is feasible to implement transform on the plane wave into the spherical wave.

\begin{equation}
e^{-ikz}=e^{-ikr\cos\theta}=\sum_{n=-\infty}^{+\infty}{c_n\left(kr\right)P_n\left(\cos\theta\right)}
\end{equation}
where the coefficient $c_n$ can be induced by $B$ function as
\begin{equation}
c_n\left(kr\right)=\frac{2l+1}{2}\int_{-1}^{1}{e^{-ikrx}P_n\left(x\right)dx=\left(2n+1\right)i^{-n}j_n(kr)\ }
\end{equation}

Therefore,

\begin{equation}
e^{-ikz}=e^{-ikr\cos\theta}=\sum_{n=-\infty}^{+\infty}{\left(2n+1\right)i^{-n}j_n(kr)P_n\left(\cos\theta\right)}
\end{equation}

\begin{figure}[htbp]
\centering
\includegraphics[scale=.65]{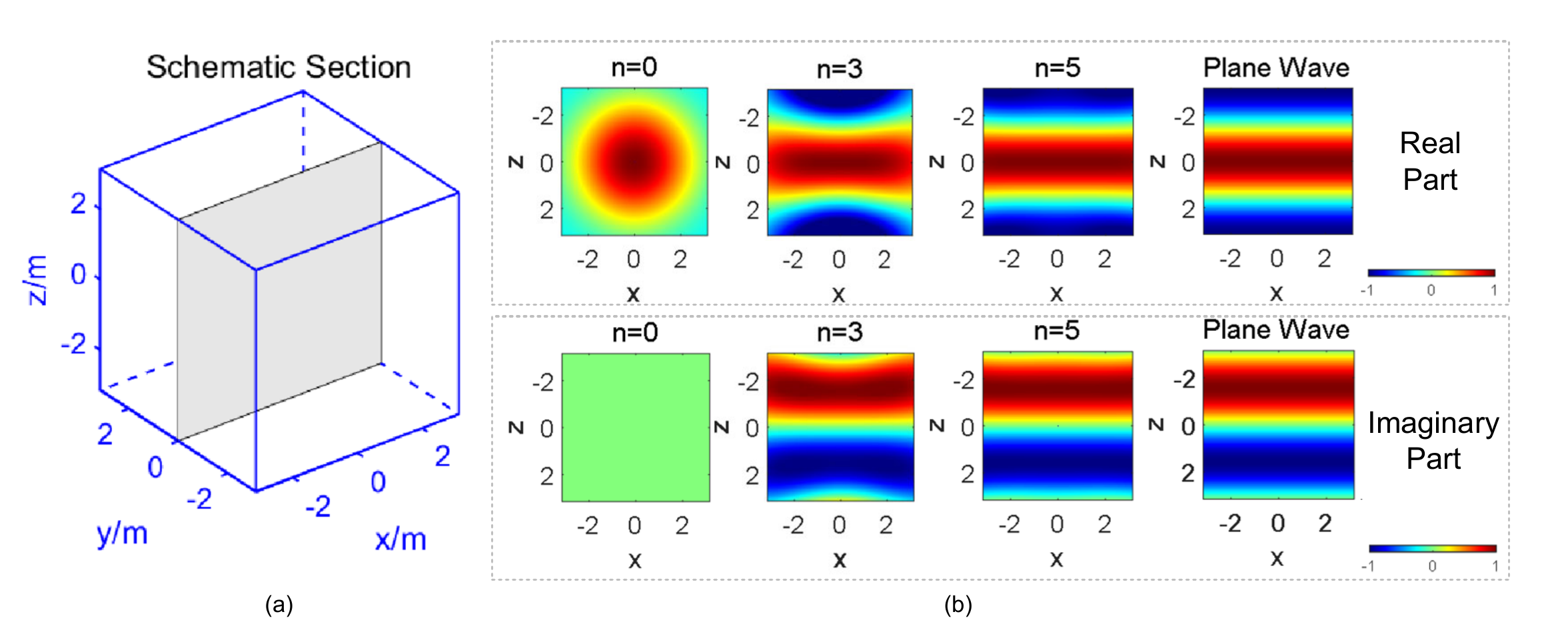}
\caption{Plane wave decompose into spherical wave. (a) Schematic section of the observation (b) Comparison of superposition of spherical wave and plane wave when the truncation n is 0,3,5 respectively.}
\label{Fig2}
\end{figure}

Fig. \ref{Fig2} (a) explains the observation section while (b) reveal the superposition of cylindrical waves and plane waves when the number of truncation terms $n$ are 1, 3 and 5 respectively.

The upper and lower groups emerge the fields of real and imaginary parts respectively. With the increase of the number of truncation terms $n$, the superposition of spherical waves is closer to the plane waves. In fact, when $n=5$, they can be approximately equal.

\subsection{Cylindrical Wave to Spherical Wave}
For a cylindrical wave, as illustrated in 2.1,  can be written as $J_n\left(k\rho\right)e^{in\phi}$, where $J_n(\cdot)$ is a Bessel function of order $n$. For spherical waves mentioned in 2.3, can be expressed as $Ac_n\left(kr\right)P_n\left(cos\theta\right)$, where $P_n\left(cos\theta\right)$ is a Legendre polynomial of order $n$. As the family of Legendre polynomials is orthogonal complete, it is feasible to implement transform on the cylindrical wave into the spherical wave. Suppose $\phi=0$ and $n=0$, we have

\begin{equation}
J_0\left(\rho\right)=J_0\left(r\sin\theta\right)=\sum_{n=0}^{\infty}{a_nj_n\left(r\right)P_n\left(\cos\theta\right)}
\end{equation}
where the coefficient $c_n$ can be induced through complex mathematical derivation. Here, the intact process is demonstrated in appendix. The result is as
\begin{equation}
J_0\left(\rho\right)=\sum_{n=0}^{\infty}\frac{\left(4n+1\right)\left(2n\right)!}{2^{2n}n!n!}j_{2n}\left(r\right)P_{2n}\left(\cos\theta\right)
\end{equation}

\begin{figure}[htbp]
\centering
\includegraphics[scale=.65]{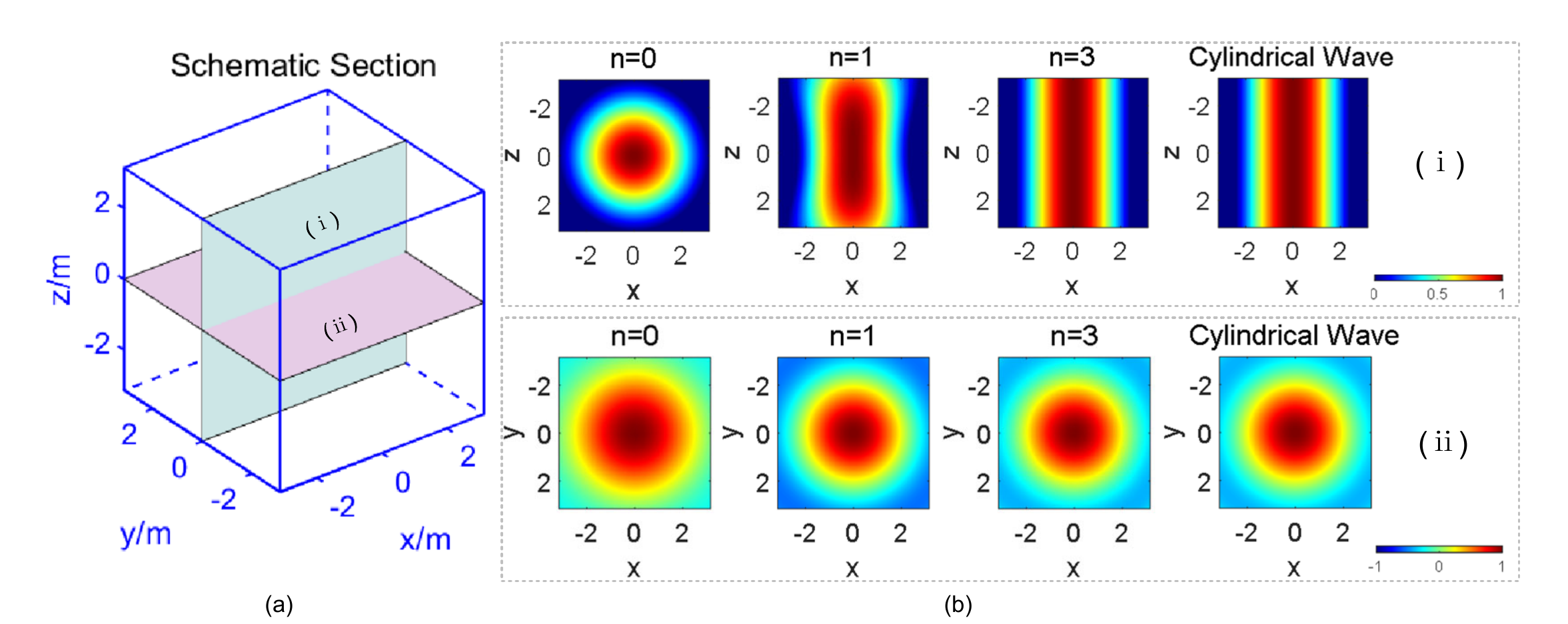}
\caption{Cylindrical wave decompose into spherical wave. (a) Schematic section of the observation (b) Comparison of superposition of spherical wave and cylindrical wave when the truncation n is 0,1,3 respectively.}
\label{Fig3}
\end{figure}

Fig. \ref{Fig3} (a) explains the observation section while (b) reveal the superposition of cylindrical waves and the cylindrical wave when the number of truncation terms $n$ are 0, 1 and 3 respectively.

The upper and lower groups emerge the fields observed at $xoy$  and $xoz$ respectively. With the increase of the number of truncation terms $n$, the superposition of spherical waves is closer to the plane waves. In fact, when $n=3$, they can be approximately equal.

\section{Application}
In the second part, we discuss the basic formula of electromagnetic wave transformation and carry out computer simulation. In this part, we will discuss some specific applications. The most classic example of electromagnetic wave transformation is the scattering of plane electromagnetic waves by objects. We mainly discuss the scattering of electromagnetic waves by ideal conductor and ideal medium.
\subsection{Perfect electric conductor}
We mainly analyze the scattering of plane waves by an infinite perfectly conducting cylinder and a perfectly conducting sphere. For an ideal conductor, it satisfies the boundary condition of zero tangential electric field, namely
\begin{equation}
E_\tau=0
\end{equation}

\subsubsection{Infinite cylinder\\}

The scattering of electromagnetic waves by an infinite perfectly conducting cylinder is actually a two-dimensional problem. According to the superposition principle, the scattering of plane wave on a metal cylinder is equal to the superposition of the scattering of infinitely many cylindrical waves. The excitation is a plane wave propagating along the $x$ direction, which can be decomposed into $TE_z$ and $TM_z$ waves. According to the relationship between the mathematical form and the physical meaning of the electromagnetic field solution, we choose the Hankel function of the second kind to represent the scattering field, and the Bessel function of the first kind to represent the incident field. The total field can be regarded as the sum of the two.

For $TM_z$ polarization, we can use only $z$-direction electric field component to represent the whole electromagnetic field.

\begin{equation}
E_z^i=E_0\sum_{n=-\infty}^{+\infty}{j^{-n}J_n\left(k\rho\right)e^{jn\phi}}
\end{equation}
\begin{equation}
E_z^s=E_0\sum_{n=-\infty}^{+\infty}{j^{-n}a_nH_n^{(2)}\left(k\rho\right)e^{jn\phi}}
\end{equation}
\begin{equation}
E_z=E_z^i+E_z^s
\end{equation}
As the PEC cylinder satisfied $E_z(a)=0$, it is feasible to configure the coefficient $a_n$. The total field can be obtained by

\begin{equation}
E_z=E_0\sum_{n=-\infty}^{+\infty}{j^{-n}\left[J_n(k\rho)-\frac{J_n(ka)}{H_n^{(2)}(ka)}H_n^{(2)}(k\rho)\right]e^{jn\varphi}}
\end{equation}

\begin{figure}[htbp]
\centering
\includegraphics[scale=.26]{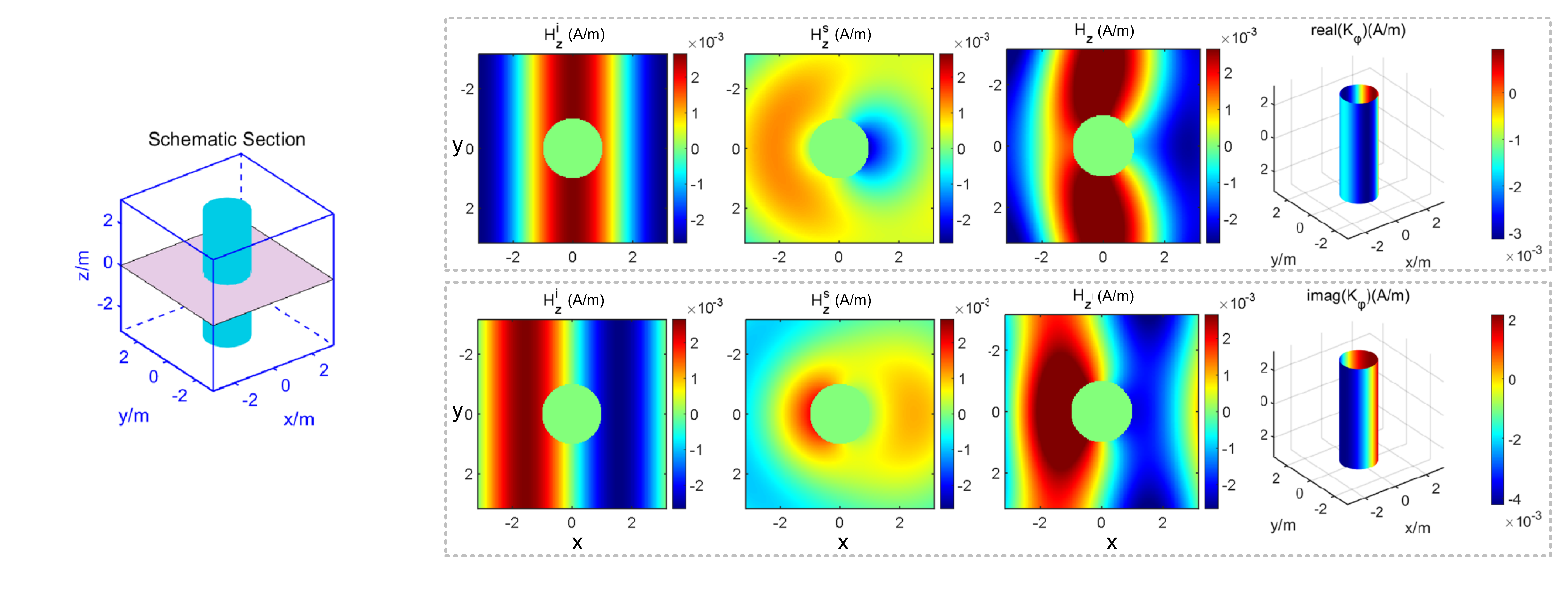}
\caption{Scattering of $TE_z$ wave by an PEC cylinder.}
\label{Fig5}
\end{figure}

For $TE_z$ polarization, we can use only $z$-direction magnetic field component to represent the whole field.
\begin{equation}
H_z^i=H_0\sum_{n=-\infty}^{+\infty}{j^{-n}J_n\left(k\rho\right)e^{jn\phi}}
\end{equation}
\begin{equation}
H_z^s=H_0\sum_{n=-\infty}^{+\infty}{j^{-n}b_nH_n^{(2)}\left(k\rho\right)e^{jn\phi}}
\end{equation}
\begin{equation}
H_z=H_z^i+H_z^s
\end{equation}
Implementing the boundary condition $E_\varphi\left(a\right)=0$, it is feasible to configure the coefficient $a_n$. The total field can be obtained by
\begin{equation}
H_z=H_0\sum_{n=-\infty}^{+\infty}{j^{-n}\left[J_n(k\rho)-\frac{J_n\prime(ka)}{H_n^{(2)}\prime(ka)}H_n^{(2)}(k\rho)\right]e^{jn\varphi}}
\end{equation}
Here, we depict the distribution of incident field, scattering field, total field and surface current in Figure. \ref{Fig5}.

\subsubsection{Sphere\\}

Here we embark on to analyze the scattering of $TEM$ waves by a PEC sphere. Considering the symmetry of the sphere, we can fix its propagation direction to $z$-axis and polarization direction to $x$-axis by selecting a reasonable coordinate system. Therefore, the incident electric field can be expressed as
\begin{equation}
E_x^i=E_0e^{-jkz}=E_0e^{-jkr\cos\theta}
\end{equation}
\begin{equation}
H_y^i=\frac{E_0}{\eta}e^{-jkz}=\frac{E_0}{\eta}e^{-jkr\cos\theta}
\end{equation}
For spherical waves, it is complicated to seek for the solution directly while a common method is to introduce the electric vector potential and magnetic vector potential. According to (6), the vector potential can be expressed as
\begin{equation}
A_r^i=E_0\frac{\cos{\varphi}}{\omega}\sum_{n=1}^{\infty}{a_n\widehat{J_n}\left(kr\right)P_n^1\left(\cos{\theta}\right)}
\end{equation}
\begin{equation}
F_r^i=\frac{E_0}{\eta}\frac{\sin{\varphi}}{\omega}\sum_{n=1}^{\infty}{a_n\widehat{J_n}\left(kr\right)P_n^1\left(\cos{\theta}\right)}
\end{equation}

Since the scattering field is an outward traveling wave, its vector potential should be $H^{(2)}_n(kr)$. Hence the total vector potentials are

\begin{equation}
A_r=E_0\frac{\cos{\varphi}}{\omega}\sum_{n=1}^{\infty}{\left[a_n\widehat{J_n}\left(kr\right)+b_n\widehat{H_n^{\left(2\right)}}\left(kr\right)\right]P_n^1\left(\cos{\theta}\right)}
\end{equation}
\begin{equation}
F_r=\frac{E_0}{\eta}\frac{\sin{\varphi}}{\omega}\sum_{n=1}^{\infty}{\left[a_n\widehat{J_n}\left(kr\right)+c_n\widehat{H_n^{\left(2\right)}}\left(kr\right)\right]P_n^1\left(\cos{\theta}\right)}
\end{equation}

Where the coefficient $a_n$, $b_n$ and $c_n$ can be obtained by the boundary condition $E_{\tau}(a)=0$,

\begin{equation}
a_n=j^{-n}\frac{\left(2n+1\right)}{n\left(n+1\right)}\\
b_n=-a_n\frac{\widehat{J_n^\prime}\left(ka\right)}{\widehat{{H_n^{\left(2\right)}}^\prime}\left(ka\right)}\\
c_n=-a_n\frac{\widehat{J_n}\left(ka\right)}{\widehat{H_n^{\left(2\right)}}\left(ka\right)}
\end{equation}

The total field can be obtained through the vector potential

\begin{equation}
E_r=\frac{1}{j\omega\mu\varepsilon}\left(\frac{\partial^2}{\partial r^2}+k^2\right)A_r
\end{equation}
\begin{equation}
E_\theta=\frac{1}{j\omega\mu\varepsilon}\frac{1}{r}\frac{\partial^2A_r}{\partial r\partial\theta}-\frac{1}{\varepsilon}\frac{1}{rsin{\theta}}\frac{\partial F_r}{\partial\varphi}
\end{equation}
\begin{equation}
E_\varphi=\frac{1}{j\omega\mu\varepsilon}\frac{1}{rsin{\theta}}\frac{\partial^2A_r}{\partial r\partial\varphi}+\frac{1}{\varepsilon}\frac{1}{r}\frac{\partial F_r}{\partial\theta}
\end{equation}

\begin{figure}[htbp]
\centering
\includegraphics[scale=.28]{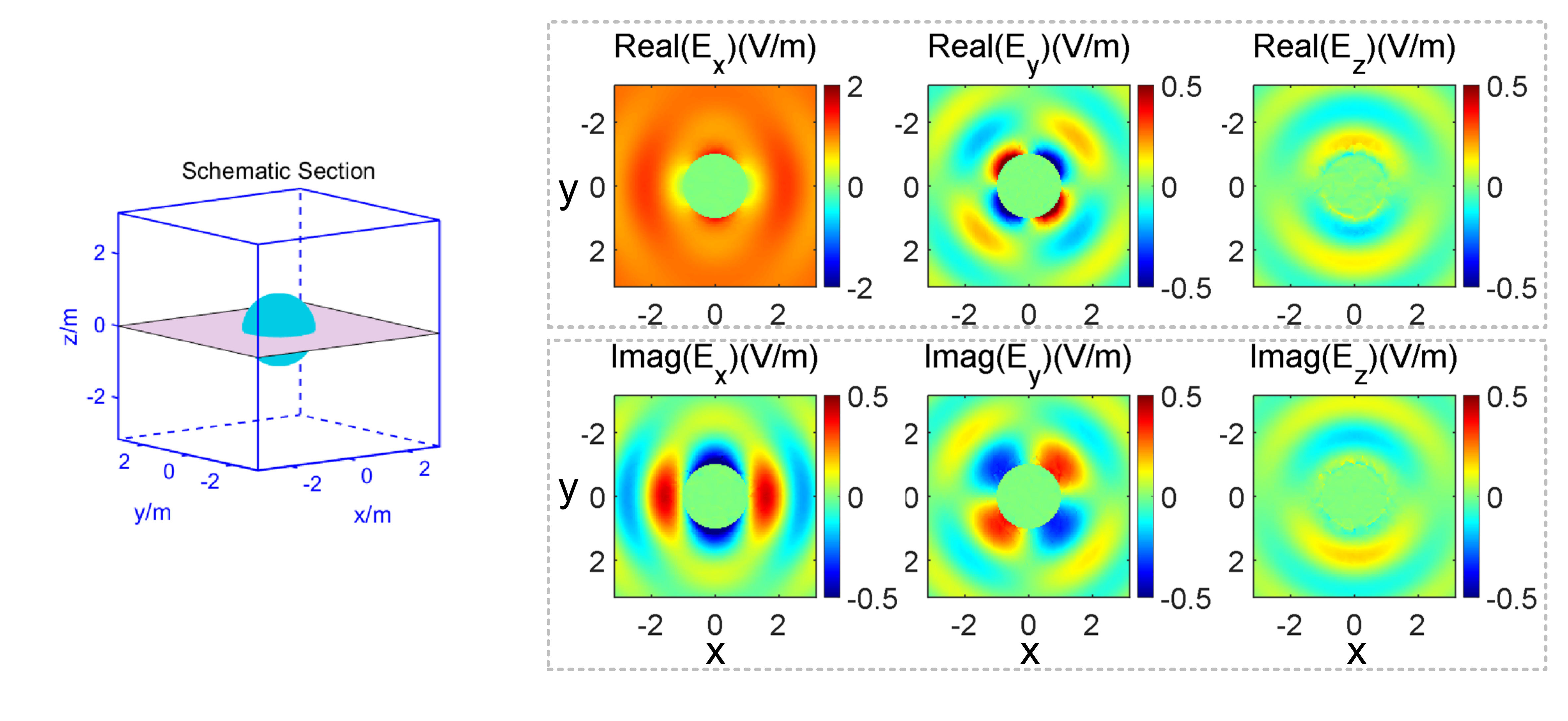}
\caption{Scattering of $TEM$ wave by an PEC cylinder.}
\label{Fig6}
\end{figure}

Through a certain coordinate transformation, we acquire the electric field distribution in rectangular coordinate system. The result is illustrated in Fig.\ref{Fig6}.

\subsection{Perfect electric dielectric}

We also analyze the scattering of plane waves by an infinite cylinder and a sphere. For perfect electric dielectrics, they satisfies the boundary condition that the tangential components of electromagnetic field are continuous, namely
\begin{equation}
E_\tau^i+E_\tau^s=E_\tau^t
\end{equation}
\begin{equation}
H_\tau^i+H_\tau^s=H_\tau^t
\end{equation}
\subsubsection{Infinite cylinder\\}
Similar to 3.1.1, we only consider $TE_z$ and $TM_z$ waves propagating along $x$ direction here. Referring to Eq. 1, the incident field and scattering field can be expressed by $J_n(\cdot)$ and $H_n^{(2)}(\cdot)$. As the transmission field is finite at the origin, it can also be expressed by $J_n(\cdot)$. Therefore, for a $TM_z$ wave propagating along the $x$ direction, we have
\begin{equation}
E_z^i=E_0\sum_{n=-\infty}^{+\infty}{j^{-n}J_n\left(k\rho\right)e^{jn\phi}}
\end{equation}
\begin{equation}
E_z^s=E_0\sum_{n=-\infty}^{+\infty}{j^{-n}a_nH_n^{(2)}\left(k\rho\right)e^{jn\phi}}
\end{equation}
\begin{equation}
E_z^t=E_0\sum_{n=-\infty}^{+\infty}{j^{-n}b_nJ_n\left(k_1\rho\right)e^{jn\phi}}
\end{equation}
The magnetic field can be derived from Faraday's law. Suppose that both media are perfect electric dielectrics, i.e. the permeability satisfies $\mu=\mu_1=\mu_0$. Utilizing the boundary condition at $\rho=a$, we have

\begin{equation}
a_n=\frac{\sqrt{\varepsilon_1}J_n\left(ka\right)J_n^\prime\left(k_1a\right)-\sqrt\varepsilon J_n\left(k_1a\right)J_n^\prime\left(ka\right)}{\sqrt\varepsilon J_n\left(k_1a\right)H_n^{\left(2\right)}\prime\left(ka\right)-\sqrt{\varepsilon_1}J_n^\prime\left(k_1a\right)H_n^{\left(2\right)}\left(ka\right)}
\end{equation}
\begin{equation}
b_n=\frac{J_n\left(ka\right)+a_nH_n^{\left(2\right)}\left(ka\right)}{J_n\left(k_1a\right)}
\end{equation}

For a $TE_z$ wave propagating along the $x$ direction, we have

\begin{equation}
H_z^i=H_0e^{-jkx}=H_0\sum_{n=-\infty}^{+\infty}{j^{-n}J_n\left(k\rho\right)e^{jn\varphi}}
\end{equation}
\begin{equation}
H_z^s=H_0\sum_{n=-\infty}^{+\infty}{j^{-n}a_nH_n^{(2)}\left(k\rho\right)e^{jn\phi}}
\end{equation}
\begin{equation}
H_z^t=H_0\sum_{n=-\infty}^{+\infty}{j^{-n}b_nJ_n\left(k_1\rho\right)e^{jn\phi}}
\end{equation}
The electric field can be derived from Ampere's law. Suppose that both media are perfect magnetic dielectrics, i.e. the permittivity satisfies $\varepsilon=\varepsilon_1=\varepsilon_0$. Utilizing the boundary condition at $\rho=a$, we have

\begin{equation}
a_n=\frac{\sqrt{\mu_1}J_n\left(ka\right)J_n^\prime\left(k_1a\right)-\sqrt\mu J_n\left(k_1a\right)J_n^\prime\left(ka\right)}{\sqrt\mu J_n\left(k_1a\right)H_n^{\left(2\right)}\prime\left(ka\right)-\sqrt{\mu_1}J_n^\prime\left(k_1a\right)H_n^{\left(2\right)}\left(ka\right)}
\end{equation}
\begin{equation}
b_n=\frac{J_n\prime\left(ka\right)+a_nH_n^{\left(2\right)}\prime\left(ka\right)}{J_n\left(k_1a\right)}
\end{equation}

\begin{figure}[htbp]
\centering
\includegraphics[scale=.26]{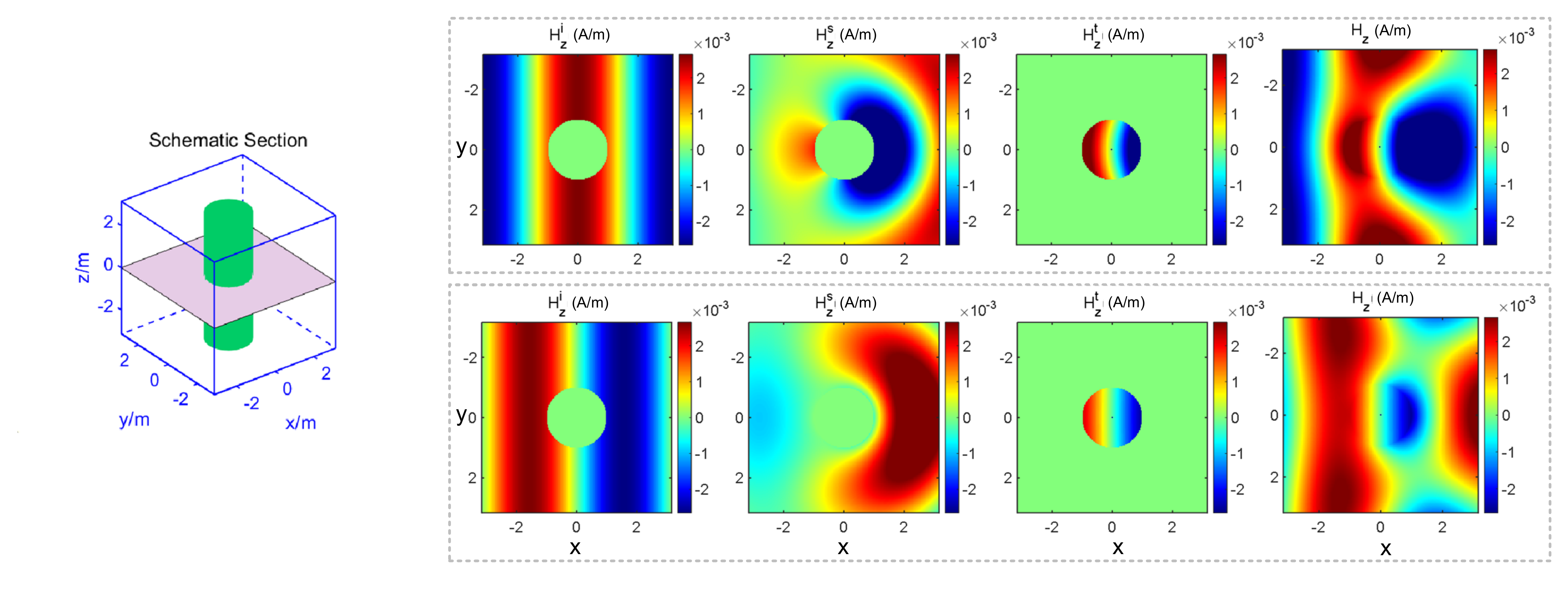}
\caption{Scattering of $TE_z$ wave by an perfect electric dielectric cylinder.}
\label{Fig8}
\end{figure}

Here, we depict the distribution of incident field, scattering field, transmission field, and total field in Figure. \ref{Fig8}.

\section{Conclusion}

This paper analyzes the common electromagnetic wave transformation problems in electromagnetics, which mainly involves the transformation between plane wave, cylindrical wave and spherical wave. Here, the corresponding mathematical expressions are derived firstly, and then the conversion process is vividly displayed by computer programs. In the application part, we analyze the scattering problem for the cylinder and the sphere with the perfect conductor and the ideal dielectric material, which proves that the transformation is able to be widely used. At the same time, this paper also corrects a mistake about wave transformation in classical textbooks, which is of certain significance to improve the research in this field.

\newpage
\appendix

\section{Proof}
Corrigendum of Eq.6-92 on page 190 in Harrington's \textit{Time-Harmonic Electromagnetic Field Theory (Second Edition)}:

In Harrington's book, the cylindrical wave $J_0\left(\rho\right)$ can be decomposed into the superposition of spherical waves as follow:
\begin{equation}
J_0\left(\rho\right)=J_0\left(r\sin\theta\right)=\sum_{n=0}^{\infty}\frac{\left(-1\right)^n\left(4n+1\right)\left(2n-1\right)!}{2^{\left(2n-1\right)}n!\left(n-1\right)!}j_{2n}\left(r\right)P_{2n}\left(\cos\theta\right)
\end{equation}
However, this is not true because the left and right sides of the equation are not equal. In fact, the correct form is as follows:
\begin{equation}
J_0\left(\rho\right)=J_0\left(r\sin\theta\right)=\sum_{n=0}^{\infty}\frac{\left(4n+1\right)\left(2n\right)!}{2^{2n}n!n!}j_{2n}\left(r\right)P_{2n}\left(\cos\theta\right)
\end{equation}
Proof

The cylindrical wave is expanded as spherical wave:
\begin{equation}
J_0\left(\rho\right)=J_0\left(r\sin\theta\right)=\sum_{n=0}^{\infty}{a_nj_n\left(r\right)P_n\left(\cos\theta\right)}
\end{equation}
As $J_0\left(r\sin\theta\right)=J_0\left(r\sin\left(\pi-\theta\right)\right)$, we have:
\begin{equation}
\sum_{n=0}^{\infty}{a_nj_n\left(r\right)P_n\left(-\cos\theta\right)}=\sum_{n=0}^{\infty}{a_nj_n\left(r\right)P_n\left(\cos\theta\right)}
\end{equation}
So
\begin{equation}
\left(1-\left(-1\right)^n\right)\sum_{n=0}^{\infty}{a_nj_n\left(r\right)P_n\left(\cos\theta\right)}=0
\end{equation}
We have $a_{2n-1}=0$:
\begin{equation}
J_0\left(r\sin\theta\right)=\sum_{n=0}^{\infty}{a_{2n}j_{2n}\left(r\right)P_{2n}\left(\cos\theta\right)}=\sum_{n=0}^{\infty}{b_nj_{2n}\left(r\right)P_{2n}\left(\cos\theta\right)}
\end{equation}
Multiply xx with $P_q\left(\cos\theta\right)\sin{\theta}$ and integrate from 0 to $\pi$, we have:
\begin{equation}
\label{cases1}
\int_{0}^{\pi}{P_q\left(\cos\theta\right)}P_{2n}\left(\cos\theta\right)\sin{\theta}d\theta=\cases{0&for $q\neq2n$\\
\frac{2}{4n+1}&for $q=2n$\\}
\end{equation}
Suppose $q=2n$, we have:
\begin{equation}
\int_{0}^{\pi}{J_0\left(r\sin{\theta}\right)}P_{2n}\left(\cos\theta\right)\sin{\theta}d\theta=\frac{2b_n}{4n+1}j_{2n}\left(r\right)
\end{equation}
As:
\begin{equation}
J_0\left(r\sin{\theta}\right)=\sum_{n=0}^{\infty}\frac{\left(-1\right)^nr^{2n}}{2^{2n}n!n!}\left(1-\cos^2{\theta}\right)^n
\end{equation}
Its $2n$ order derivative to $r$ is:
\begin{equation}
J_0^{\left(2n\right)}\left(r\sin{\theta}\right)=\frac{\left(-1\right)^n\left(2n\right)!}{2^{2n}n!n!}\left(1-\cos^2{\theta}\right)^n+M\left(r\right)
\end{equation}
Where $M(r)$ is the polynomial of R higher than degree 1. Suppose $r=0$, the $2n$ order derivative of the left side of the equation to $r$ is:
\begin{equation}
L=\int_{0}^{\pi}{\frac{\left(-1\right)^n\left(2n\right)!}{2^{2n}n!n!}\left(1-\cos^2{\theta}\right)^n}P_{2n}\left(\cos\theta\right)\sin{\theta}d\theta
\end{equation}
For $l < 2n$, Legendre polynomials satisfy the following equation:
\begin{equation}
\label{cases2}
\int_{-1}^{1}x^lP_{2n}\left(x\right)dx=\cases{0&for $l<2n$\\
2^{2n+1}\frac{\left(2n\right)!\left(2n\right)!}{\left(4n+1\right)!}&for $l=2n$\\}
\end{equation}
Therefore the left side can be obtained:
\begin{equation}
L=\frac{\left(-1\right)^n\left(2n\right)!}{2^{2n}n!n!}\cdot2^{2n+1}\frac{\left(-1\right)^n\left(2n\right)!\left(2n\right)!}{\left(4n+1\right)!}=2\frac{\left(2n\right)!\left(2n\right)!\left(2n\right)!}{n!n!\left(4n+1\right)!}
\end{equation}
For the right side, as:
\begin{equation}
j_{2n}\left(r\right)=\frac{\sqrt\pi}{2}\sum_{l=0}^{\infty}\frac{\left(-1\right)^l}{l!\Gamma\left(l+2n+\frac{3}{2}\right)}\left(\frac{r}{2}\right)^{4l+2n}
\end{equation}
Its $2n$ order derivative to $r$ is:
\begin{equation}
j_{2n}^{\left(2n\right)}\left(r\right)=\frac{\sqrt\pi}{2}\frac{1}{2^{2n}}\frac{\left(2n\right)!}{\Gamma\left(2n+\frac{3}{2}\right)}+N\left(r\right)
\end{equation}
Where $N(r)$ is the polynomial of R higher than degree 1. Suppose $r=0$, the $2n$ order derivative of the left side of the equation to $r$ is:
\begin{equation}
   R=\frac{\sqrt\pi}{2}\frac{1}{2^{2n}}\frac{\left(2n\right)!}{\Gamma\left(2n+\frac{3}{2}\right)}\cdot\frac{2b_n}{4n+1} 
\end{equation}
Using $\mathrm{\Gamma}\left(z+1\right)=z\mathrm{\Gamma}\left(z\right)$, we can get:
\begin{equation}
\Gamma\left(2n+\frac{3}{2}\right)=\frac{4n+1}{2}\cdot\frac{4n-1}{2}\cdots\frac{1}{2}\mathrm{\Gamma}\left(\frac{1}{2}\right)=\frac{\left(4n+1\right)!!}{2^{2n+1}}\mathrm{\Gamma}\left(\frac{1}{2}\right)
\end{equation}
Considerating that $\mathrm{\Gamma}\left(\frac{1}{2}\right)=\sqrt\pi$, we have:
\begin{equation}
\Gamma\left(2n+\frac{3}{2}\right)=\frac{\sqrt\pi\left(4n+1\right)!}{2^{2n+1}\cdot2^{2n}\left(2n\right)!}
\end{equation}
Therefore:
\begin{equation}
    R=\frac{b_n2^{2n+1}\left(2n\right)!\left(2n\right)!}{\left(4n+1\right)\left(4n+1\right)!}
\end{equation}
As $L=R$, we have:
\begin{equation}
b_n=\frac{\left(4n+1\right)\left(2n\right)!}{2^{2n}n!n!}
\end{equation}
To sum up,
\begin{equation}
    J_0\left(\rho\right)=J_0\left(r\sin\theta\right)=\sum_{n=0}^{\infty}\frac{\left(4n+1\right)\left(2n\right)!}{2^{2n}n!n!}j_{2n}\left(r\right)P_{2n}\left(\cos\theta\right)
\end{equation}

\end{document}